\documentclass[3p,sort&compress,twocolumn]{elsarticle}
\usepackage{atlasphysics}
\usepackage{subfigure}
\usepackage{multirow}
\usepackage{amssymb}
\usepackage{hyperref}
\newcommand\htollll{$H\to ZZ^{(*)}\to\ell^+\ell^-\ell^{+}\ell^{-}$}
\newcommand\htollllp{$H\to ZZ^{(*)}\to\ell^+\ell^-\ell^{'+}\ell^{'-}$}
\newcommand\htollllbrief{$H\to ZZ^{(*)}\to 4\ell$}

\newcommand\progname[1]{{\sc #1}}

\newcommand{\tabscript}[3]{%
 	\setlength{\fboxrule}{0pt}%
 	\fbox{\ensuremath{#1^{#2}_{#3}}}%
}%for super and sub scripts  in tabulars
\newcommand\lumiAverage{\ensuremath{2.1~\ifb}}
\newcommand\lumiFourMuon{\ensuremath{2.28~\ifb}}
\newcommand\lumiFourElectron{\ensuremath{1.98~\ifb}}
\newcommand\lumiTwoMuonTwoElectron{\ensuremath{1.96~\ifb}}
\newcommand\sensitivityExpected{1.01}
\newcommand\sensitivityObserved{0.99}
\newcommand\sensitivityMass{\ensuremath{194\:\gev}}
\newcommand\candtotal{27~}
\newcommand\candmu{12~}
\newcommand\candmix{9~}
\newcommand\cande{6~}
\newcommand\bkgexp{$24\pm 4$~}

\journal{Physics Letters B} 
\begin{document}
\begin{frontmatter}
\title{\vspace{-0.5cm} \flushright{\normalsize{CERN-PH-EP-2011-144}} \\ \center{Search for the Standard Model Higgs boson in the decay channel \htollllbrief{} with the ATLAS detector}}
\author{The ATLAS Collaboration}
\address{}

\begin{abstract}
A search for the Standard Model Higgs boson in the decay channel \htollll{}, where $\ell=e,\mu$, is presented. 
Proton-proton collision data at $\sqrt{s}=7\:\tev$ recorded with the ATLAS detector and corresponding to an average integrated luminosity of \lumiAverage~are compared to the Standard Model expectations. Upper limits on the production cross section of a Standard Model Higgs boson with a mass between 110 and $600\:\gev$ are derived. 
The observed (expected) 95\% confidence level upper limit on the production cross section for a Higgs boson with a mass of \sensitivityMass, the region with the best expected sensitivity for this search, is \sensitivityObserved~(\sensitivityExpected) times the Standard Model prediction. The Standard Model Higgs boson is excluded at 95\% confidence level in the mass ranges $191-197$, $199-200$ and $214-224\:\gev$.
\end{abstract}

\begin{keyword}
LHC, ATLAS, Higgs, leptons
\end{keyword}

\end{frontmatter}

\section{Introduction}

The search for the Standard Model (SM) Higgs boson~\cite{Englert:1964et,Higgs:1964pj,Guralnik:1964eu} is a major goal of the Large Hadron Collider (LHC) programme. Direct searches at the CERN LEP $e^+e^-$ collider led to a lower limit on the Higgs boson mass, $m_H$, of $114.4\:\gev$ at 95\% confidence level (CL)~\cite{Barate:2003sz}. The searches at the Fermilab Tevatron $p\bar{p}$ collider have excluded at 95\% CL the region $156\:\gev < m_H < 177\:\gev$~\cite{tevhiggs}. Results from the 2010 LHC run extended the search in the region $200\:\gev< m_H<600\:\gev$ by excluding a Higgs boson with cross section larger than $5-20$ times the SM prediction~\cite{atlas2010,cms2010}.

This letter presents a search for the SM Higgs boson in the mass range from 110 to $600\:\gev$ in the channel \htollllp{}, where $\ell,\ell^{'}=e,\mu$. 
Three distinct final states, \mmmm~($4\mu$), \eemm~($2e2\mu$), and \eeee~($4e$), are selected. 
The largest background to this search comes from continuum $ZZ^{(*)}$ production. For $m_H<180\:\gev$, contributions from $Z+\rm{jets}$ and $t\bar{t}$ processes, where the additional charged leptons arise either from semi-leptonic decays of heavy flavour or from light flavour jets misidentified as leptons, are important.
The $pp$ collision data were recorded with the ATLAS detector at the LHC at $\sqrt{s}=7\:\tev$ and correspond to an average integrated luminosity of \lumiAverage~\cite{lumi2011}. 

\section{The ATLAS Detector}
The ATLAS detector \cite{Aad:2008zzm} is a multi-purpose particle physics apparatus with forward-backward symmetric cylindrical geometry\footnote{ATLAS uses a right-handed coordinate system with its origin at the nominal interaction point. The z-axis is along the beam pipe, the x-axis points to the centre of the LHC ring and the y-axis points upward. Cylindrical coordinates (r,$\phi$) are used in the transverse plane, $\phi$ being the azimuthal angle around the beam pipe. The pseudorapidity $\eta$ is defined as $\eta=-\ln[\tan(\theta/2)]$ where $\theta$ is the polar angle.}. The inner tracking detector (ID) consists of a silicon pixel detector, a silicon microstrip detector, and a transition radiation tracker. The ID is surrounded by a thin superconducting solenoid providing a $2\:\rm{T}$ magnetic field. A high-granularity lead-liquid argon (LAr) sampling calorimeter measures the energy and the position of electromagnetic showers. An iron-scintillator tile calorimeter provides hadronic coverage in the central rapidity range. The end-cap and forward rapidity regions are instrumented with LAr calorimetry for both electromagnetic and hadronic measurements. The muon spectrometer (MS) surrounds the calorimeters and consists of three large superconducting toroids, each with eight coils, a system of precision tracking chambers, and detectors for triggering. A three-level trigger system selects events to be recorded for offline analysis. 

\section{Data and Simulation Samples}
The accumulated data are subjected to quality requirements ensuring that the relevant detector components were operating normally. The resulting average integrated luminosity of \lumiAverage\ corresponds to \lumiFourMuon{}, \lumiTwoMuonTwoElectron{} and \lumiFourElectron{} for the $4\mu$, $2e2\mu$ and $4e$~final states, respectively.

The \htollllbrief{} signal is modelled using the \progname{powheg} Monte Carlo (MC) event generator~\cite{powheg4,powheg5}, which calculates separately the gluon and vector-boson fusion production mechanisms with matrix elements up to next-to-leading order (NLO). The Higgs boson transverse momentum, \pt, spectrum is reweighted to the calculation of Ref.~\cite{Bozzi:2005wk}, providing QCD corrections up to next-to-leading order and QCD soft-gluon resummations up next-to-next-to-leading log (NNLL). \progname{powheg} is interfaced to \progname{pythia}~\cite{pythia} for showering and hadronization, which in turn is interfaced to \progname{photos}~\cite{Golonka:2005pn} for QED radiative corrections in the final state and to \progname{tauola}~\cite{Jadach:1993hs,Golonka:2003xt} for the simulation of $\tau$ decays. 

The cross sections for Higgs~boson production, the corresponding branching fractions, as well as their uncertainties~\cite{LHCHiggsCrossSectionWorkingGroup:2011ti}, are derived to next-to-next-to-leading order (NNLO) in QCD 
for the gluon fusion~\cite{Harlander:2002wh,Anastasiou:2002yz,Ravindran:2003um,Anastasiou:2008tj,deFlorian:2009hc,Baglio:2010ae} and vector boson fusion~\cite{Bolzoni:2010xr} processes.
In addition, QCD soft-gluon resummations up to NNLL are  available for the 
gluon fusion process~\cite{Catani:2003zt}, while the NLO electroweak (EW) corrections are applied to both the gluon fusion~\cite{Aglietti:2004nj,Actis:2008ug} and
vector boson fusion~\cite{Ciccolini:2007jr,Ciccolini:2007ec} processes. The Higgs boson decay branching ratio to the four-lepton final state is predicted by \progname{prophecy4f}~\cite{Bredenstein:2006rh,Bredenstein:2006ha}, which includes the complete NLO QCD+EW corrections, interference effects between identical final state fermions and leading two-loop heavy Higgs boson corrections to the four-fermion width. Table~\ref{tab:MCSignal} gives the production cross sections for the $H \rightarrow 4\ell$ for several Higgs boson masses.   
\begin{table}[ht]
\begin{center}
\caption{Higgs boson production cross sections for both gluon and vector-boson fusion processes in $pp$ collisions at $\sqrt{s}=7\:{\rm TeV}$. The cross sections include the branching ratio of $H \rightarrow 4\ell$, with $\ell = e, \mu$. The errors are the total theoretical systematic uncertainty.\label{tab:MCSignal}}
\vspace{0.1cm}
{\footnotesize
\begin{tabular}{cccc}
\hline\hline
$m_{H}$ & $\sigma\left(gg\to H\right)$ & $\sigma\left(qq\to H\right)$ & ${\rm BR}\left(H\to 4\ell\right)$ \\
$[{\rm GeV}]$ & $[{\rm pb}]$ & $[{\rm pb}]$ & $\cdot 10^{-3}$\\
\hline
130\rule[-1mm]{0mm}{4.7mm} & $14.1^{+2.7}_{-2.1}$& $1.154^{+0.032}_{-0.027}$ & 0.19\\ 
150\rule[-1mm]{0mm}{4.7mm} & $10.5^{+2.0}_{-1.6}$& $0.962^{+0.028}_{-0.021}$ & 0.38\\ 
200\rule[-1mm]{0mm}{4.7mm} & $5.2^{+0.9}_{-0.8}$& $0.637^{+0.022}_{-0.015}$ & 1.15\\
240\rule[-1mm]{0mm}{4.7mm} & $3.6\pm 0.6$& $0.464^{+0.018}_{-0.012}$ & 1.32\\
300\rule[-1mm]{0mm}{4.7mm} & $2.4\pm 0.3$& $0.301^{+0.014}_{-0.008}$ & 1.38\\
400\rule[-1mm]{0mm}{4.7mm} & $2.0\pm 0.3$& $0.162^{+0.010}_{-0.005}$ & 1.21\\
600\rule[-1mm]{0mm}{4.7mm} & $0.33\pm0.06$& $0.058^{+0.005}_{-0.002}$ & 1.23\\
\hline\hline
\end{tabular}
}
\end{center}
\end{table}

The $ZZ^{(*)}$ background is generated using \progname{pythia}, taking into account $Z-\gamma$ interference. For the inclusive total cross section and the shape of the $m_{ZZ^{(*)}}$ spectrum, the \progname{mcfm}~\cite{Campbell:1999ah,Campbell:2011bn} prediction is used, which includes both quark-antiquark annihilation at QCD NLO and gluon fusion. The inclusive $Z$ boson production, $Z+\rm{jets}$, is modelled using \progname{alpgen}~\cite{Mangano:2002ea} and is divided into $Z$+light flavour jets and $Zb\bar{b}$; overlaps between the two samples are removed. Specifically, $b\bar{b}$ pairs with separation $\Delta R=\sqrt{\Delta \phi^2+\Delta\eta^2}\geq 0.4$ between the $b$-jets are taken from the matrix-element calculation, whereas for $\Delta R<0.4$ the parton-shower jets are taken. \progname{pythia} is also used as a cross-check of the \progname{alpgen} results. In this search the $Z+\rm{jets}$ production is normalized from the data, but for comparisons the QCD NNLO \progname{fewz}~\cite{Melnikov:2006kv,Anastasiou:2003ds} and the \progname{mcfm}~\cite{Campbell:1999ah,Campbell:2011bn} cross section calculations are used for the inclusive $Z$ boson and the $Zb\bar{b}$ production, respectively. 
The $t\bar{t}$ background is modelled using \progname{mc@nlo}~\cite{Frixione:2003ei} and is normalized to the approximately NNLO cross section calculated using \progname{hathor}~\cite{Aliev:2010zk}. Both \progname{alpgen} and \progname{mc@nlo} are interfaced to \progname{herwig}~\cite{herwig} for parton shower hadronization and to \progname{jimmy}~\cite{jimmy} for the underlying event simulations. 
\par
All generated events undergo a full detector simulation performed using GEANT4~\cite{GEANT4,simuAtlas}.

The number of $pp$ interactions in the same bunch crossing (pileup) is included in the simulation. The MC samples are reweighted to reproduce the observed distribution in the data. 

\section{Physics Object Identification and Event Selection}
The data considered in this analysis were selected using single-lepton triggers. For electrons the threshold on the transverse energy, $E_{T}$, was $20-22\:\gev$ depending on the LHC instantaneous luminosity and for muons the threshold on $\pt$ was $18\:\gev$. Both triggers are more than 99.5\% efficient for events passing the offline selection described below.

Electron candidates consist of clusters of energy deposited in the electromagnetic calorimeter associated to ID tracks. The electrons must satisfy the ``medium'' electron criteria~\cite{Aad:2010yt}, which require the shower profiles to be consistent with those expected for electromagnetic showers and a well reconstructed ID track pointing to the corresponding cluster. The electron transverse momentum is computed from the cluster energy and the track direction.  

Muon candidates are reconstructed by matching ID tracks with either full or partial tracks in the MS~\cite{Aad:2010yt}. 
For the former case, the two independent momentum measurements are combined, whereas for the latter case the momentum is measured using the ID information only, with the MS providing muon identification. To reject cosmic rays, tracks are required to be consistent with having originated from the primary vertex, defined as the reconstructed vertex with the highest $\sum\pt^2$ of associated tracks.

Leptons from Higgs boson decays are expected to be isolated and to originate from a common vertex. Track and calorimeter isolation as well as transverse impact parameter significance requirements are therefore applied to further reduce the $Z+\rm{jets}$ and $t\bar{t}$ contributions. The sum of \pt~of tracks within $\Delta R<0.2$ of the lepton divided by the lepton \pt~is required to be less than 0.15, while the sum \et~of the calorimeter cells within $\Delta R<0.2$ around the lepton divided by the lepton \pt~is required to be less than 0.3. In the case of electrons, the calorimeter cells corresponding to the electromagnetic shower are subtracted. The transverse impact parameter significance, defined as the transverse impact parameter of the lepton with respect to the primary vertex divided by its uncertainty, for the two lowest \pt~leptons of the quadruplet in events with $m_{4\ell}<190\:\gev$ is required to be less than 3.5 and 6 for muons and electrons respectively. The selection efficiency of the isolation and impact parameter requirements has been studied using data both for isolated leptons, with $Z\rightarrow\ell\ell$ decays  and non-isolated leptons from semi-leptonic $b$ and $c-$quark decays in a heavy-flavour enriched dijet sample. Good agreement is observed between data and simulation.

Higgs boson candidates are searched by selecting two same-flavour, opposite-sign isolated lepton pairs in an event. Each lepton must satisfy $\pt>7\:\gev$ and be measured in the pseudorapidity range $\left|\eta \right|<2.47$ for electrons and $\left|\eta\right|<2.5$ for muons. The electron \pt~threshold is increased to $15\:\gev$ in the transition region between the barrel and end-cap calorimeters ($1.37< |\eta| <1.52$). At least two leptons must have $\pt >20\:\gev$. The leptons are required to be well separated from each other with $\Delta R>0.1$. The invariant mass of the lepton pair closest to the nominal $Z$ boson mass ($m_{Z}$) is denoted by $m_{12}$ and it is required that $\left|m_{Z}-m_{12}\right|<15\:\gev$. The invariant mass of the remaining lepton pair, $m_{34}$, is required to be lower than $115\:\gev$ and greater than a threshold depending on the reconstructed four lepton mass, $m_{4\ell}$, as summarized in Table~\ref{tab:kincuts}. The final discriminating variable is $m_{4\ell}$, where the Higgs boson production would appear as a clustering of events. The width of the reconstructed Higgs boson mass distribution is dominated by experimental resolution at low $m_{H}$ values, with a full-width at half-maximum (FWHM) which varies according to decay mode and is between 4.5 ($4\mu$) and 6.5 ($4e$) GeV for $m_{H} = 130\:\gev$. At high $m_{H}$ the reconstructed width is dominated by the natural width of the Higgs boson with a FWHM of approximately $35\:\gev$ at $m_{H} = 400\:\gev$.
\begin{table*}[!htb]
  \begin{center}
    \caption{Thresholds applied to $m_{34}$ for reference values of $m_{4\ell}$ (see text). For other $m_{4\ell}$ values, the selection requirement is obtained via linear interpolation. \label{tab:kincuts}}
  \vspace{0.1cm}
  \begin{tabular}{cccccccccc}
    \hline\hline
  $\tabscript{m}{}{4\ell}$ (\gev) & $\leq$120 & 130 & 140 & 150 & 160 & 165 & 180 & 190 & $\geq$200 \\
  threshold  (\gev) & 15   & 20  & 25 & 30 & 30 & 35 & 40 & 50 & 60 \\
  \hline\hline
  \end{tabular}
  \end{center}
\end{table*}

\begin{table*}[!htb]
\begin{center}
\caption{The expected numbers of background events, with their systematic uncertainty, separated into ``Low mass'' ($m_{4\ell}<180\:\gev$) and ``High mass'' ($m_{4\ell}\geq 180\:\gev$) regions. The expected numbers of signal events for different $m_{H}$ hypotheses and the observed numbers of events are also presented.\label{tab:EventYields}}
\begin{tabular}{ccccccc}
\hline\hline
 & \multicolumn{2}{c}{\mmmm}& \multicolumn{2}{c}{\eemm} & \multicolumn{2}{c}{\eeee} \\
 & Low mass & High mass& Low mass & High mass& Low mass & High mass\\
\hline
Integrated Luminosity\rule[-1mm]{0mm}{4.7mm} & \multicolumn{2}{c}{ \lumiFourMuon{}}& \multicolumn{2}{c}{\lumiTwoMuonTwoElectron{}} & \multicolumn{2}{c}{\lumiFourElectron{}} \\
\hline
$\tabscript{ZZ}{(*)}{}$             & 1.02$\pm$0.15  & 7.7$\pm$1.2   & 0.99$\pm$0.16 & 9.6$\pm$1.4   & 0.39$\pm$0.09 & 3.6$\pm$0.5 \\
$Z$, $Zb\bar{b}$, $t\bar{t}$        & 0.06$\pm$0.01  & 0.01$\pm$0.01 & 0.29$\pm$0.11 & 0.15$\pm$0.06  & 0.23$\pm$0.09 & 0.12$\pm$0.05 \\
\hline
Total Background                    & 1.08$\pm$0.15  & 7.7$\pm$1.2   & 1.28$\pm$0.19 & 9.8$\pm$1.4   & 0.62$\pm$0.13 & 3.7$\pm$0.5\\
\hline 
Data & 1 & 11 & 1& 8 & 1& 5\\
\hline 
$m_{H}=130\:\gev$ & 0.42 $\pm$ 0.07 && 0.40 $\pm$ 0.06 && 0.14 $\pm$ 0.03 &\\
$m_{H}=150\:\gev$ & 0.98 $\pm$ 0.15 && 0.97 $\pm$ 0.15 && 0.34 $\pm$ 0.06 &\\
$m_{H}=200\:\gev$ && 2.26 $\pm$ 0.33 && 2.64 $\pm$ 0.38 && 0.98 $\pm$ 0.14 \\
$m_{H}=240\:\gev$ && 1.74 $\pm$ 0.25 && 2.24 $\pm$ 0.32 && 0.88 $\pm$ 0.13 \\
$m_{H}=300\:\gev$ && 1.18 $\pm$ 0.17 && 1.64 $\pm$ 0.23 && 0.64 $\pm$ 0.09 \\
$m_{H}=400\:\gev$ && 0.86 $\pm$ 0.13 && 1.23 $\pm$ 0.18 && 0.52 $\pm$ 0.08 \\
$m_{H}=600\:\gev$ && 0.15 $\pm$ 0.02 && 0.23 $\pm$ 0.04 && 0.10 $\pm$ 0.02 \\
\hline\hline
\end{tabular}
\end{center}
\end{table*}
\begin{figure*}[!htb]
  \centering 
  \subfigure[\label{fig:sub_m12data}]{\includegraphics[width=0.32\linewidth]{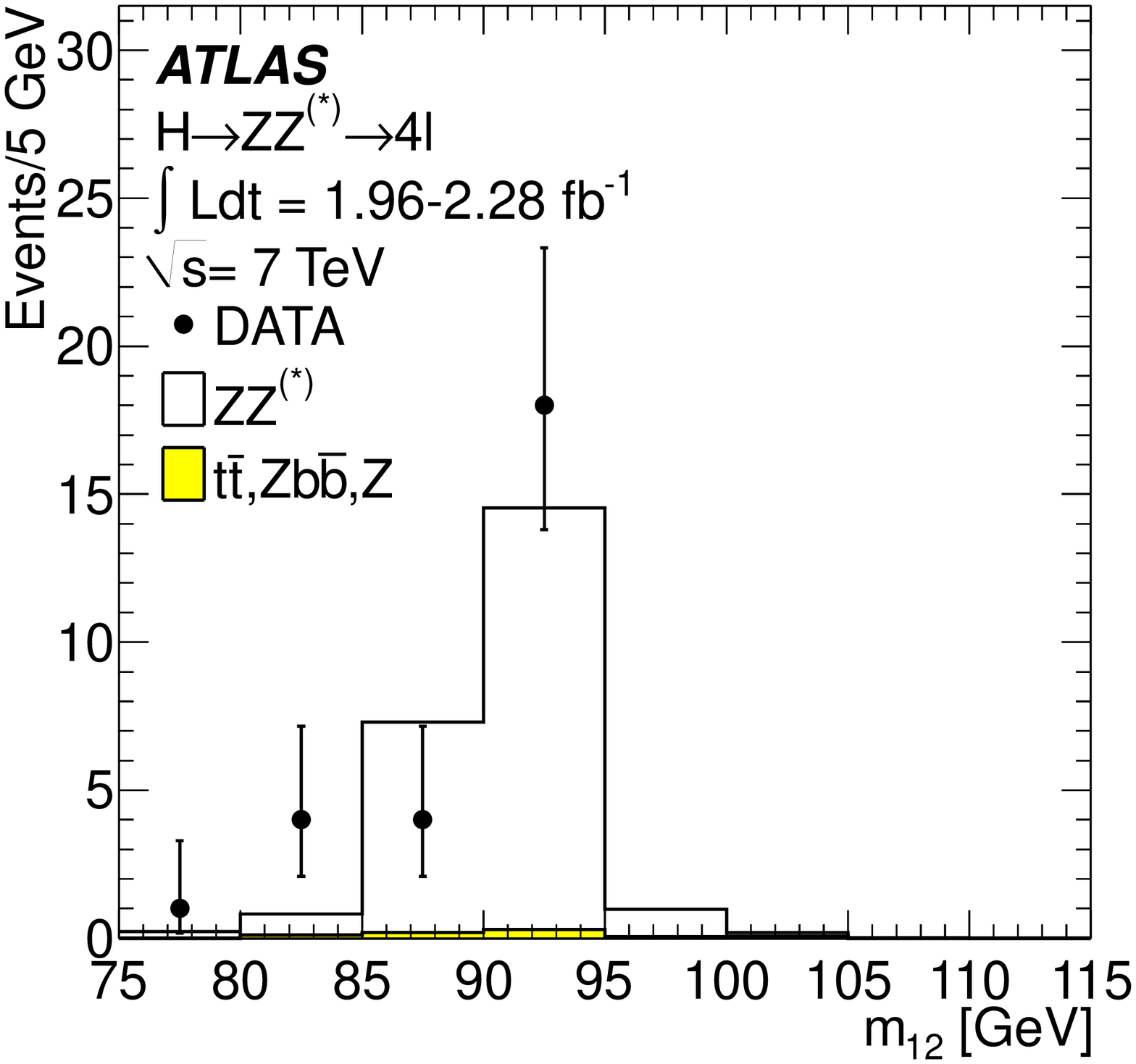}}
  \subfigure[\label{fig:sub_m34data}]{\includegraphics[width=0.32\linewidth]{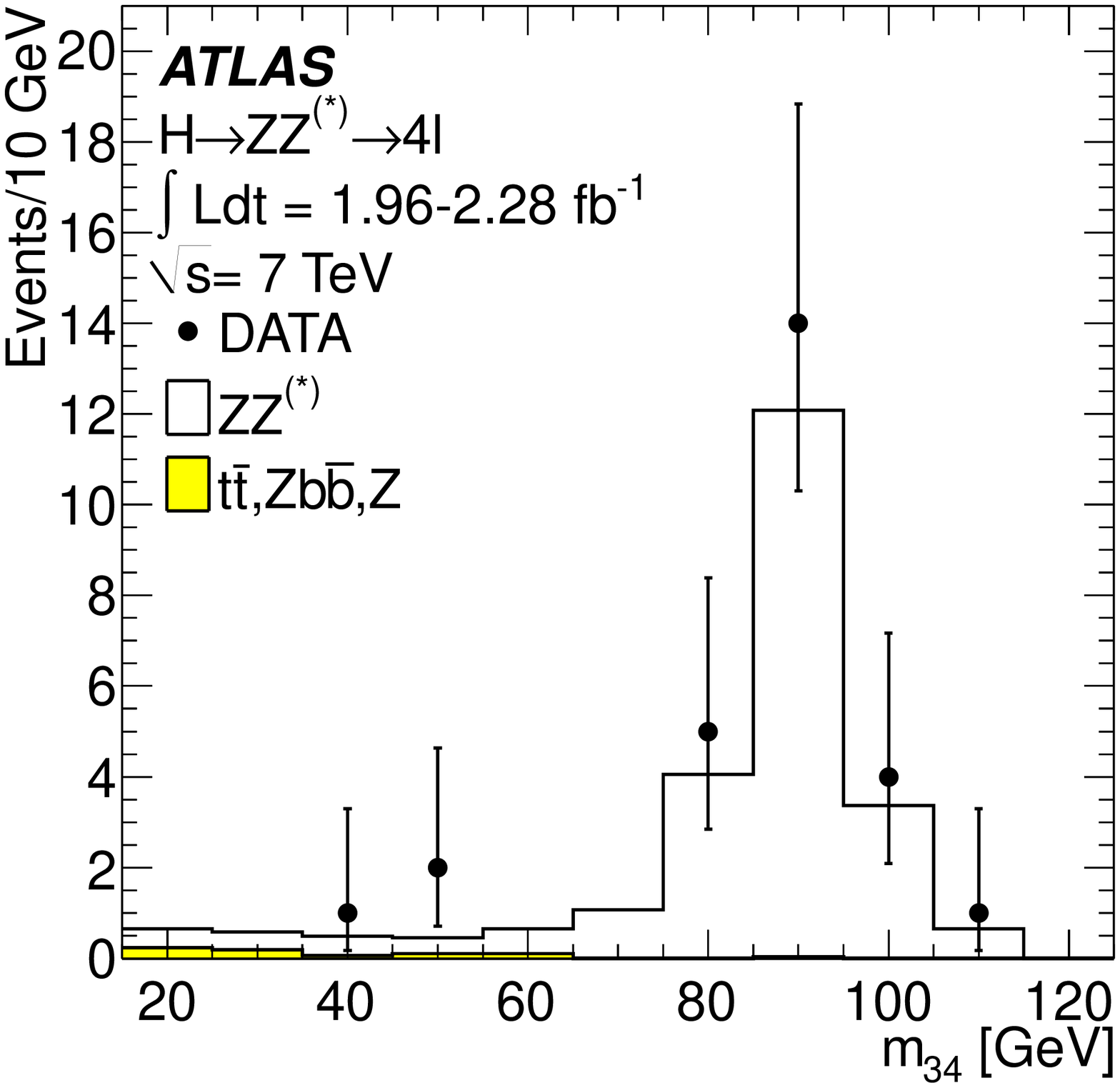}}
  \subfigure[\label{fig:sub_m4ldata}]{\includegraphics[width=0.32\linewidth]{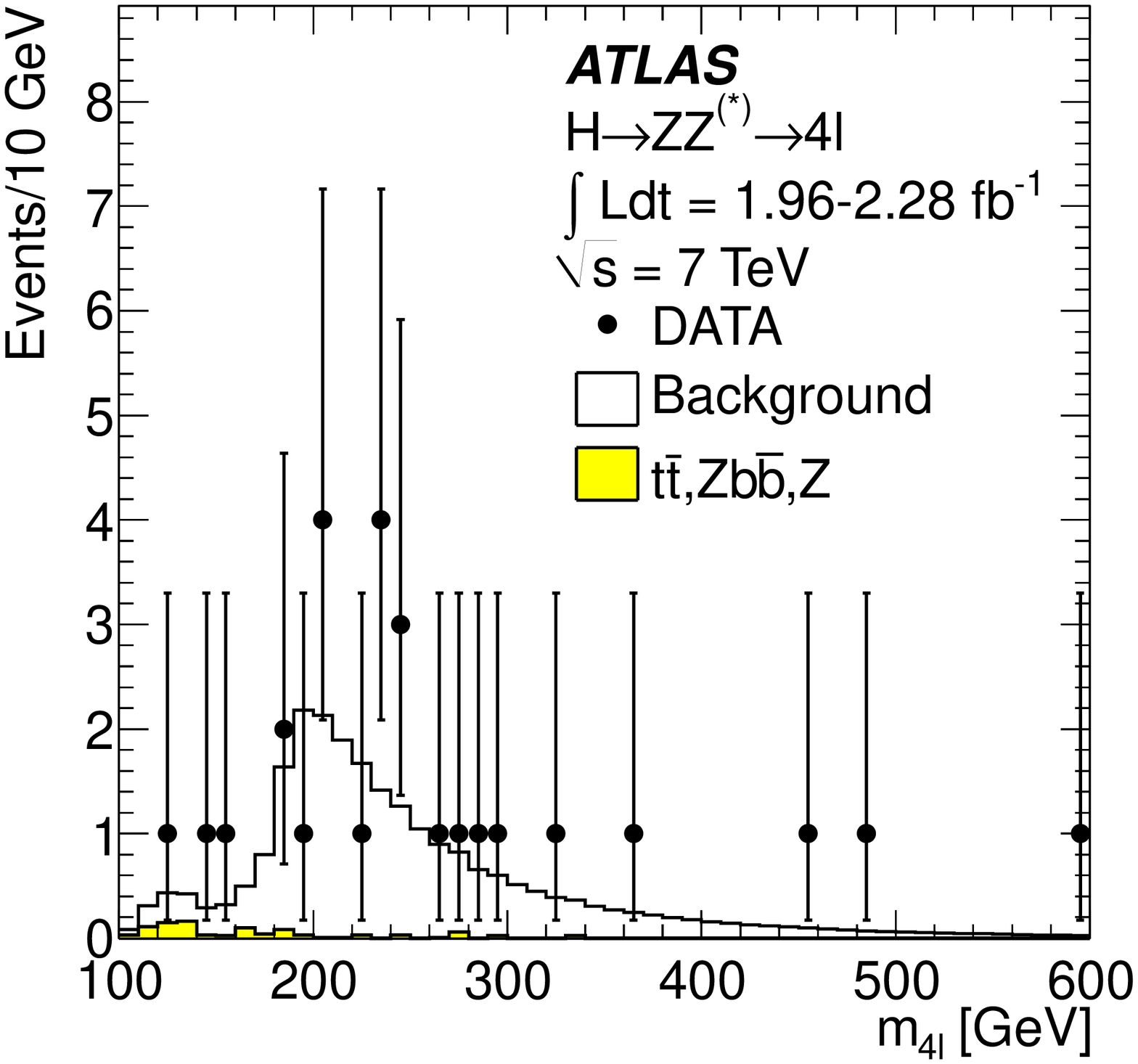}}
\caption{Invariant mass distributions \subref{fig:sub_m12data} $m_{12}$, \subref{fig:sub_m34data} $m_{34}$, and \subref{fig:sub_m4ldata} $m_{4\ell}$ for the selected candidates. The data (dots) are compared to the background expectations from the dominant $ZZ^{(*)}$ process and the sum of $t\bar{t}$, $Zb\bar{b}$ and $Z$+light flavour jets processes. Error bars represent 68.3\% central confidence intervals.\label{fig:finalMasses}}
\end{figure*}

\section{Background Estimation}
The dominant $ZZ^{(*)}$ background is estimated using MC simulation. Generated events are required to pass the complete analysis selection and the final yield is normalized to the integrated luminosity. 

The $t\bar{t}$ background is also estimated using MC simulation. Comparison of data to MC predictions, in a control sample of events with opposite sign electron-muon pairs consistent with the $Z$ boson mass and with one or two additional charged leptons, are used to verify that the $t\bar{t}$ background is small with respect to the dominant $ZZ^{(*)}$ process and in agreement with expectation.

The $Z+\rm{jets}$ background is normalized using data. The control sample is formed by selecting events with a pair of same-flavour, opposite-sign isolated leptons  consistent with the $Z$ boson mass, $\left |m_{Z}-m_{12}\right|<15\:\gev$, and a second same-flavour, opposite-sign lepton pair where only kinematic, but no isolation or impact parameter, requirements are applied. 
At this stage, the dominant background source depends on the flavour of the second lepton pair: $Z$+light flavour jets dominates the final states with a second electron pair, while $Zb\bar{b}$ production dominates the final states with a second muon pair after the contributions from $t\bar{t}$, $ZZ^{(*)}$, and muons from in-flight $\pi$ and $K$ decays which correspond to 44\% of the event yield are subtracted. The observed background, which is found to be in good agreement with expectation, is extrapolated to the signal region by means of the MC simulation. 

\section{Systematic Uncertainties}
Uncertainties on lepton reconstruction and identification efficiency, and on the momentum resolution and momentum scale are determined using samples of $W$, $Z$ and $J/\psi$ decays. The muon efficiency uncertainty results in an acceptance uncertainty on the signal and the irreducible background which is uniform over the mass range of interest and amounts to 1.7\% (1.2\%) for the $4\mu$ ($2e2\mu$) channel. The uncertainty on the electron efficiency results in an acceptance uncertainty of 3\% (2\%) for the $4e$ ($2e2\mu$) channel at $m_{4\ell}=600\:\gev$ reaching 15\% (6\%) at $m_{4\ell}=110\:\gev$.

A conservative theoretical uncertainty of 15\% is assigned to the $ZZ^{(*)}$ background contribution~\cite{Campbell:2009kg}. The $Z$+light flavour jets and $Zb\bar{b}$ backgrounds are evaluated using data. A systematic uncertainty between 20\% and 40\% is assigned on their normalization to account for the statistical uncertainty in the control sample and the MC-based extrapolation to the signal region. The uncertainty on the $\ttbar$ cross section is found to be 10\% by adding linearly the contributions from variations of the renormalization and factorization scales to those of the parton distribution functions.

The theoretical uncertainties on the Higgs~boson production cross section are 15--20\% for the gluon fusion process and 3--9\% for the vector-boson fusion process~\cite{LHCHiggsCrossSectionWorkingGroup:2011ti}, depending on the Higgs~boson mass\footnote{The limits presented in this study for $m_{H}>200\:\gev$ assume cross sections based on on-shell Higgs boson production and decay and use MC generators with an ad-hoc Breit-Wigner Higgs line shape. Recently potentially important effects related to off-shell Higgs boson production and interference effects between the Higgs boson signal and backgrounds have been discussed~\cite{LHCHiggsCrossSectionWorkingGroup:2011ti,Anastasiou:2011pi}. The inclusion of such effects may affect limits at high Higgs masses ($m_H>400\:\gev$).}. They include uncertainties on the QCD scale and on the parton distribution functions~\cite{Botje:2011sn,Martin:2009iq,Lai:2010vv,Ball:2011mu}. An additional 2\% uncertainty is added to the signal selection efficiency due to the modelling of the signal kinematics. This is evaluated by comparing signal samples generated with \progname{pythia} and the default \progname{powheg} samples. 

The overall uncertainty on the total integrated luminosity is 3.7\%~\cite{lumi2011}.

\section{Results}
\begin{figure}[ht]
  \centering 
  \includegraphics[width=0.90\linewidth]{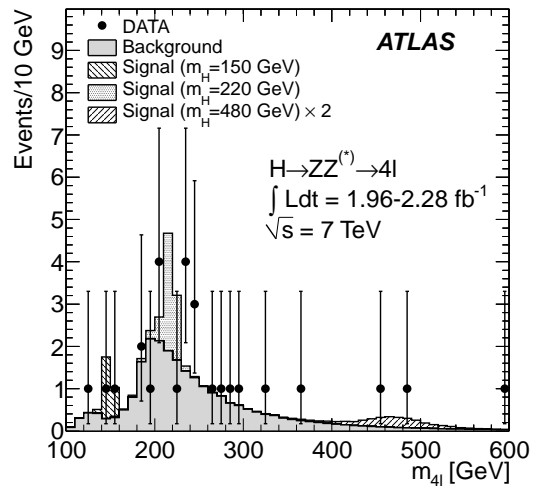}
  \caption{$m_{4\ell}$ distribution of the selected candidates, compared to the background expectation. Error bars represent 68.3\% central confidence intervals.The signal expectation for three $m_{H}$ hypotheses is also shown.\label{fig:finalMassesSignal}}
\end{figure}

The number of events observed in each final state, separately for $m_{4\ell}<180\:\gev$ and $m_{4\ell}\geq 180\:\gev$, are compared with the expectations for background and signal for various $m_{H}$ hypotheses in Table~\ref{tab:EventYields}. In total \candtotal candidate events are selected by the analysis: \candmu $4\mu$, \candmix $2e2\mu$, and \cande $4e$ events, while in the same mass range \bkgexp events are expected from the background processes. The $m_{12}$, $m_{34}$, and $m_{4\ell}$ mass spectra are shown in Fig.~\ref{fig:finalMasses}. The $m_{4\ell}$ distribution for the total background and several signal hypotheses is compared to the data in Fig.~\ref{fig:finalMassesSignal}. The selected events have been examined visually and no evidence for reconstruction problems was identified.

\begin{figure}[ht]
\centering  
\includegraphics[width=0.97\linewidth]{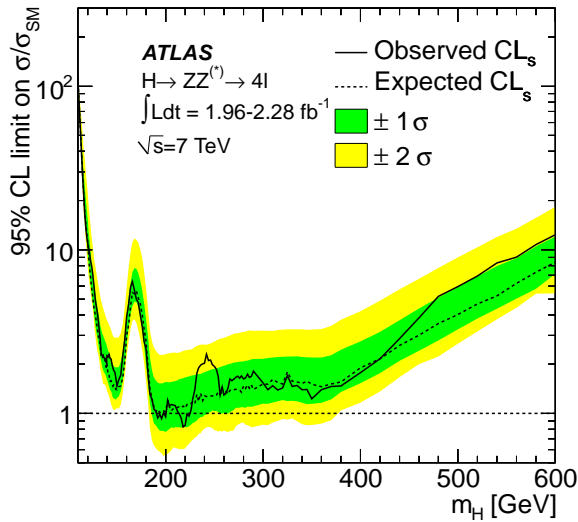}
\caption{The expected (dashed) and observed (full line) 95\% CL upper limits on the Higgs boson production cross section as a function of the Higgs boson mass, divided by the expected SM Higgs boson cross section. The green and yellow bands indicate the expected sensitivity with $\pm 1\sigma$ and $\pm 2\sigma$ fluctuations, respectively.\label{fig:exclusion}}
\end{figure}

Upper limits are set on the Higgs boson cross section at 95\% CL, using the $CL_s$ modified frequentist formalism~\cite{cls} with the profile likelihood test statistic~\cite{Cowan:2010js}. The test statistic is evaluated with a maximum likelihood fit of signal and background models to the observed $m_{4\ell}$ distribution. Figure~\ref{fig:exclusion} shows the expected and observed 95\% CL cross section upper limits as a function of $m_{H}$ and Table~\ref{tab:limit} summarizes the numerical values  for selected $m_{H}$ points. The consistency with the background-only hypothesis is quantified using the $p$-value, the probability that a background-only experiment fluctuates more than the observation. The most significant deviation from the background-only hypothesis is observed for $m_{H}=242\:\gev$ with a $p$-value of 4.9\%. These results do not account for the so-called ``look-elsewhere'' effect~\cite{Gross:2010qma}.  The SM Higgs boson is excluded at 95\% CL in the mass ranges $191-197$, $199-200$ and $214-224\:\gev$.


\begin{table}[ht]
\begin{center}
\caption{Median expected and observed 95\% CL upper limits on the Higgs boson production cross section for several Higgs boson masses, divided by the expected SM Higgs boson cross section.\label{tab:limit}}
\vspace{0.1cm}
\begin{tabular}{cr@{.}lr@{.}l}
\hline\hline
Mass (\gev) & \multicolumn{2}{r}{Expected} & \multicolumn{2}{r}{Observed} \\
\hline
130   &  3&29 & 4&11 \\
150   &  1&39 & 1&47 \\
200   &  1&03 & 0&96 \\
240   &  1&28 & 2&03 \\
300   &  1&51 & 1&54\\
400   &  1&91 & 1&77 \\
600   &  8&40 & 12&34 \\
\hline\hline
\end{tabular}
\end{center}
\end{table}

\section{Summary}
A search for the Standard Model Higgs boson in the decay channel \htollllbrief~based on \lumiAverage~ of data recorded by the ATLAS detector at $\sqrt{s}=7\:\tev$ during the 2011 run, has been presented. No significant excess of candidates is observed in the mass range between 110 and $600\:\gev$ with respect to the expected SM background. The observed (expected) 95\% CL upper limit on the Higgs boson production cross section, in units of the SM cross section, is \sensitivityObserved~(\sensitivityExpected) for $m_H=\sensitivityMass$, the region with the best expected sensitivity for this search. The SM Higgs boson is excluded at 95\% CL in the mass ranges $191-197$, $199-200$ and $214-224\:\gev$.

\section{Acknowledgements}

We thank CERN for the very successful operation of the LHC, as well as the
support staff from our institutions without whom ATLAS could not be
operated efficiently.

We acknowledge the support of ANPCyT, Argentina; YerPhI, Armenia; ARC,
Australia; BMWF, Austria; ANAS, Azerbaijan; SSTC, Belarus; CNPq and FAPESP,
Brazil; NSERC, NRC and CFI, Canada; CERN; CONICYT, Chile; CAS, MOST and
NSFC, China; COLCIENCIAS, Colombia; MSMT CR, MPO CR and VSC CR, Czech
Republic; DNRF, DNSRC and Lundbeck Foundation, Denmark; ARTEMIS, European
Union; IN2P3-CNRS, CEA-DSM/IRFU, France; GNAS, Georgia; BMBF, DFG, HGF, MPG
and AvH Foundation, Germany; GSRT, Greece; ISF, MINERVA, GIF, DIP and
Benoziyo Center, Israel; INFN, Italy; MEXT and JSPS, Japan; CNRST, Morocco;
FOM and NWO, Netherlands; RCN, Norway; MNiSW, Poland; GRICES and FCT,
Portugal; MERYS (MECTS), Romania; MES of Russia and ROSATOM, Russian
Federation; JINR; MSTD, Serbia; MSSR, Slovakia; ARRS and MVZT, Slovenia;
DST/NRF, South Africa; MICINN, Spain; SRC and Wallenberg Foundation,
Sweden; SER, SNSF and Cantons of Bern and Geneva, Switzerland; NSC, Taiwan;
TAEK, Turkey; STFC, the Royal Society and Leverhulme Trust, United Kingdom;
DOE and NSF, United States of America.

The crucial computing support from all WLCG partners is acknowledged
gratefully, in particular from CERN and the ATLAS Tier-1 facilities at
TRIUMF (Canada), NDGF (Denmark, Norway, Sweden), CC-IN2P3 (France),
KIT/GridKA (Germany), INFN-CNAF (Italy), NL-T1 (Netherlands), PIC (Spain),
ASGC (Taiwan), RAL (UK) and BNL (USA) and in the Tier-2 facilities
worldwide.

\bibliographystyle{elsarticle-num}
\bibliography{H4lPaper}

\clearpage
\onecolumn
% ATLAS Collaboration author list for 11-JUL-2011
% Data extracted on 12-Aug-2011 for paperid 119
%\documentclass[11pt]{article}
%\usepackage{a4wide}\begin{document}
\begin{flushleft}
{\Large The ATLAS Collaboration}

\bigskip

G.~Aad$^{\rm 48}$,
B.~Abbott$^{\rm 111}$,
J.~Abdallah$^{\rm 11}$,
A.A.~Abdelalim$^{\rm 49}$,
A.~Abdesselam$^{\rm 118}$,
O.~Abdinov$^{\rm 10}$,
B.~Abi$^{\rm 112}$,
M.~Abolins$^{\rm 88}$,
H.~Abramowicz$^{\rm 153}$,
H.~Abreu$^{\rm 115}$
% ...
\bigskip

% ...
$^{10}$ Institute of Physics, Azerbaijan Academy of Sciences, Baku, Azerbaijan\\
$^{11}$ Institut de F\'isica d'Altes Energies and Departament de F\'isica de la Universitat Aut\`onoma  de Barcelona and ICREA, Barcelona, Spain\\
% ...
$^{48}$ Fakult\"{a}t f\"{u}r Mathematik und Physik, Albert-Ludwigs-Universit\"{a}t, Freiburg i.Br., Germany\\
$^{49}$ Section de Physique, Universit\'e de Gen\`eve, Geneva, Switzerland\\
% ...
$^{88}$ Department of Physics and Astronomy, Michigan State University, East Lansing MI, United States of America\\
% ...
$^{111}$ Homer L. Dodge Department of Physics and Astronomy, University of Oklahoma, Norman OK, United States of America\\
$^{112}$ Department of Physics, Oklahoma State University, Stillwater OK, United States of America\\
% ...
$^{115}$ LAL, Univ. Paris-Sud and CNRS/IN2P3, Orsay, France\\
% ...
$^{118}$ Department of Physics, Oxford University, Oxford, United Kingdom\\
% ...
$^{153}$ Raymond and Beverly Sackler School of Physics and Astronomy, Tel Aviv University, Tel Aviv, Israel\\
% ...
\end{flushleft}

%\end{document}
\end{document}
\bye